\documentclass[11pt,onecolumn,oneside]{osajnl}

\journal{jocn} 

\title{Secrecy Performance of $ \alpha-\kappa-\mu $ Shadowed Fading Channel}

\author[1]{A. S. M. Badrudduza}
\author[2]{S. H. Islam}
\author[3]{M. K. Kundu}
\author[4]{I. S. Ansari}

\affil[1]{Department of Electronics \& Telecommunication Engineering, Rajshahi University of Engineering \& Technology (RUET), Rajshahi-6204, Bangladesh}
\affil[2]{Department of Electrical \& Electronic Engineering, RUET}
\affil[3]{Department of Electrical \& Computer Engineering, RUET}
\affil[3]{James Watt School of Engineering, University of Glasgow, Glasgow G12 8QQ, United Kingdom}

\begin{abstract}
In this paper, the physical layer security aspects of a wireless framework over $\alpha-\kappa-\mu$ shadowed (AKMS) fading channel are examined by acquiring closed-form novel expressions of average secrecy capacity, secure outage probability (SOP), and strictly positive secrecy capacity. The lower bound of SOP is derived along with the asymptotic expression of SOP at the high signal-to-noise ratio regime in order to achieve secrecy diversity gain. Capitalizing on these expressions, the consequences due to the simultaneous occurrence of fading and shadowing are quantified. Finally, Monte-Carlo simulations are demonstrated to assess the correctness of the expressions.

\quad

\end{abstract}

\begin{document}

\maketitle

\section{Introduction}
Multipath fading and shadowing are two common effects in practical wireless scenarios that are liable for the degradation of propagated wireless signals. In particular, shadowing characterizes long-term variation of the signals whereas multipath fading arises due to the interference between multiple delayed versions of the transmitted signal. It is noteworthy that in realistic wireless applications, both parameters affect the secrecy performance remarkably.

In recent times, the authors are showing their intense interest in the physical layer security (PLS) issue that utilizes the time-varying property of fading channels to enhance the information security \cite{lei2015performance,lei2017secrecy,bhargav2016secrecy,srinivasan2018secrecy,8556482, 8543055} rather than utilizing the classical cryptography approaches.
Secrecy performance over Generalized-$K$ (GK) fading channel was analyzed in \cite{lei2015performance} in terms of average secrecy capacity (ASC), secure outage probability (SOP), and strictly positive secrecy capacity (SPSC). In \cite{lei2017secrecy}, authors exhibited superiority of $\alpha-\mu$ fading channel by stating that the secrecy performance over GK fading channel can be easily approximated via $\alpha-\mu$ fading channel. As generalized channels exhibit precedence over multipath fading channels, security over $\kappa-\mu$ and $\kappa-\mu$ shadowed fading channel was analyzed in \cite{bhargav2016secrecy,srinivasan2018secrecy} showing some classical models as special cases.
The natural generalization of $\alpha-\mu$ and $\kappa-\mu$ fading channels was performed in \cite{8556482} by modelling a secure scenario over $\alpha-\kappa-\mu$ and $\alpha-\eta-\mu$ fading channels which are further generalized by $\alpha-\eta-\kappa-\mu$ fading channel in \cite{8543055}. The authors examined the secrecy characteristics and showed that for high average signal-to-noise ratio (SNR) of main and eavesdropper channels, an ASC ceiling is attained.

The aforecited works in \cite{lei2015performance,lei2017secrecy,bhargav2016secrecy,8556482,8543055} exhibit the impact of fading during the analysis of PLS. The impact of shadowing in PLS was presented only in \cite{srinivasan2018secrecy}. But practical scenarios experience fading and shadowing simultaneously, hence the research gaps in these existing works are addressed in this paper via tackling the impact of both (fading and shadowing) on secrecy performance over $\alpha-\kappa-\mu$ shadowed (AKMS) fading model.
In brief, the contributions are:

\begin{itemize}
    \item The existing models in \cite{lei2015performance,lei2017secrecy,bhargav2016secrecy,srinivasan2018secrecy,8556482} can be obtained as a special case of the proposed model in this work. Moreover, this model unifies the performance evaluation of all classical multipath models such as Nakagami-$m$, Rician, Nakagami-$q$, Weibull, etc. \cite[Table~I,]{noor2021on}.
    \item Secrecy performance evaluation is accomplished by deducing analytical expressions for SPSC, SOP, and ASC in closed-form. Additionally, asymptotic outage characteristics at high SNR regime are also demonstrated. The correctness of the deduced expressions is analyzed via Monte-Carlo simulations.
\end{itemize}

The rest of this paper is structured as follows: System model is illustrated in Section-II whereas the channel modeling is performed in Section-III. The formulation of the performance metrics is introduced in Section-IV. Numerical results are demonstrated in Section-V and finally, the conclusions are discussed in Section-VI.

\section{System Model}

In the proposed model, a source, $\mathcal{T}$, transmits sensitive messages to an authorized receiver, $\mathcal{H}$, via main ($\mathcal{T}-\mathcal{H}$) link. A passive eavesdropper, $\mathcal{K}$, also exists in that network that is trying to overhear the secret transmission between $\mathcal{T}$ and $\mathcal{H}$ via the eavesdropper ($\mathcal{T}-\mathcal{K}$) link. All nodes are assumed to have a single antenna. Assuming $\mathcal{T}-\mathcal{H}$ and $\mathcal{T}-\mathcal{K}$ links experience severe fading and shadowing concurrently, those are modeled as independently and identically distributed AKMS fading channels. The channel gain between $\mathcal{T}$ and $\mathcal{H}$ is denoted as $f_{th}\in \mathbb{C}^{1 \times 1}$. Similarly, the channel coefficient for $\mathcal{T}-\mathcal{K}$ link is denoted as $l_{tk}\in \mathbb{C}^{1 \times 1}$.
Considering $P_{t}$, $P_{h}$, and $P_{k}$ represent transmit power from $\mathcal{T}$, noise power at $\mathcal{H}$, and noise power at $\mathcal{K}$, respectively,
the instantaneous SNRs of $\mathcal{T}-\mathcal{H}$ and $\mathcal{T}-\mathcal{K}$ links are given by $\gamma_{h}=\frac{P_{t}}{P_{h}}\|f_{th}\|^{2}$ and $\gamma_{k}=\frac{P_{t}}{P_{k}}\|l_{tk}\|^{2}$, respectively. The information in $\mathcal{T}-\mathcal{H}$ link is secure if $\mathcal{T}$ transmits messages at secrecy rate i.e. a rate at which eavesdroppers are incapable of wiretapping \cite{wyner1975wire}.
All the system parameters are summarized in Table \ref{Notation}.
\begin{table}[ht]
\centering
\caption{Notations and their descriptions}
\label{Notation}
\begin{tabular}{ c c }
\hline
Notations & Descriptions \\
\hline
$\alpha_{i}$, $\kappa_{i}$, $\mu_{i}$, $m_{i}$ &  Non-negative real shape parameters
\\
${_1}F_1(.)$  & Confluent hypergeometric function
\\
${_2}F_1(.)$ & Gauss hypergeometric function
\\
$\Gamma(.)$ & Gamma operator
\\
$\gamma(.,.)$ & Lower incomplete gamma function
\\
$\mathcal{R}_{t}$ & Target secrecy rate
\\ \hline
\end{tabular}
\end{table}  

\section{Channel Model}

The PDF of instantaneous SNR is given by \cite[Eq.~(2),]{noor2021on}
\begin{align}
  \label{abcd6}
  f_{i}(\gamma)  = a_{i} \gamma^{\tilde{\alpha}_{i}\mu_{i}-1} e^{- b_{i} \gamma^{\tilde{\alpha}_{i}}} {_1}F_1(m_{i},\mu_{i};d_{i}\gamma^{\tilde{\alpha}_{i}}),
\end{align}
where $i \in (h,k)$, $c_{i}= \biggl(\frac{(\mu_{i}\kappa_{i}+m_{i})^{m_{i}}\Gamma(\mu_{i})}{m_i^{m_i} \Gamma(\mu_i+\tilde{\alpha}_{i}^{-1}) {_2}F_1(m_i,\mu_i+\tilde{\alpha}_{i}^{-1};\mu_i;\frac{\mu_i \kappa_i}{\mu_i \kappa_i+m_i})}\biggl)^{\tilde{\alpha}_{i}}$, $\tilde{\alpha}_{i}=\frac{\alpha_{i}}{2}$, $ a_{i} = \frac{{m_{i}}^{m_{i}} \alpha_{i}  }{2 {c_{i}}^{\mu_{i}} \Gamma(\mu_{i}) (\mu_{i} \kappa_{i}+m_{i})^{m_{i}}\overline{\gamma}_{i}^{\tilde{\alpha}_{i}} \mu_{i}}$,   $b_{i} =\frac{1}{c_{i} \overline{\gamma}_{i}^{\tilde{\alpha}_{i}}} $,  $d_{i}=\frac{\mu_{i} \kappa_{i}}{c_{i} (\mu_{i} \kappa_{i} + m_{i})\overline{\gamma}_{i}^{\tilde{\alpha}_{i}}}$,
and $\overline{\gamma}_{i}$ represents the average SNR of the channels. Utilizing \cite[Eq.~(9.14.1),]{gradshteyn2014table},  \eqref{abcd6} is simplified as
\begin{align}
\label{abcd13}
f_{i}(\gamma)  =a_i \sum_{j_i =0}^{\infty} \frac{\mathcal{A}_i d_i^{j_i}}{j_i !} \gamma^{\tilde{\alpha}_{i}(\mu_i + j_i)-1} e^{-b_i \gamma^{\tilde{\alpha}_{i}}},
\end{align}
where $\mathcal{A}_i=\frac{\Gamma(\mu_i) \Gamma(m_i+j_i)}{\Gamma(m_i) \Gamma(\mu_i+j_i)}$. Integrating \eqref{abcd13} with respect to $\gamma$ by making use of \cite[Eqs.~(3.381.8) and (8.352.6),]{gradshteyn2014table}, lower incomplete gamma function can modified and the CDF of $\gamma$ can be derived as
\begin{align}
\label{abcd1}
    F_i(\gamma)=  1-\frac{ a_i}{\tilde{\alpha}_i}\sum_{j_i=0}^{\infty}\mathcal{B}_i e^{-b_i \gamma^{\tilde{\alpha}_i}}\sum_{l_i=0}^{\mu_i+j_i-1}\frac{{b_i}^{l_i} \gamma^{\tilde{\alpha}_i l_i }}{l_i!},
\end{align}
where $\mathcal{B}_i = \frac{\mathcal{A}_i {d_i}^{j_i}  (\mu_i+j_i-1)!}{  {b_i}^{\mu_i+j_i}  j_i!}$.

\section{Performance Metrics}

\subsection{Average Secrecy Capacity (ASC) Performance}
ASC can be expressed as \cite{lei2015performance}
\begin{align}
    \label{asc1}
    \overline{C_s}= \Im_1+\Im_2-\Im_3,
\end{align}
where the expressions of $\Im_1$, $\Im_2$, and $\Im_3$ are derived in Appendix \ref{ASC}.

\subsection{Secrecy Outage Probability (SOP) Performance}
The SOP refers to the probability that $\mathcal{S}_{c}$ falls below $\mathcal{R}_{t}$ \cite{badrudduza2020enhancing}. Mathematically,
the lower bound of SOP is given by 
\begin{align}
\label{abcd3}
SOP_{L}=\int_{0}^{\infty}F_{h}(\varphi\gamma_{k})f_{k}(\gamma_{k})d\gamma_{k}.
\end{align}

\subsubsection{Exact Analysis}

By substituting \eqref{abcd6} and \eqref{abcd1} into \eqref{abcd3}, the exact expression of lower bound of the SOP is derived as
\begin{align}
\label{abcd4}
SOP_{L,e}=&1- \frac{ a_h a_k}{\tilde{\alpha}^2}\sum_{j_h=0}^{\infty}  \sum_{l_h=0}^{\mu_h+j_h-1} \frac{ \mathcal{B}_h\varphi^{\tilde{\alpha}l_h} }{ b_h^{ -l_h} l_h!}  
\sum_{j_k=0}^{\infty} 
\frac{\mathcal{A}_k d_k^{j_k}(\mu_k+j_k+l_h-1)!}{j_k!(b_k+b_h \varphi^{\tilde{\alpha}})^{\mu_k+j_k+l_h}}.
\end{align}

\subsubsection{Asymptotic Analysis}
In order to achieve more insights of various system parameters on the system's outage behaviour, SOP analysis at high SNR regime are shown. The asymptotic expression for lower bound of the SOP is obtained as
\begin{align}
\label{abcd5}
SOP_{L,a}&=\frac{ a'_h a'_k}{\tilde{\alpha}^2 \mu_h} \varphi^{\alpha \mu_h}  \sum_{j_k=0}^{\infty} \frac{\mathcal{A}_k (d'_k)^{j_k} }{j_k! } \biggl( \frac{\overline{\gamma}_k}{\overline{\gamma}_h} \biggl)^{\alpha \mu_h}
\frac{(\mu_h+\mu_k+j_k-1)!}{(b'_k)^{\mu_h+\mu_k+j_k}}.
\end{align}
The proof is shown in Appendix \ref{SOP}. From \eqref{abcd5}, it is observed that diversity gain of the system is $\mathcal{G} = \tilde{\alpha} \mu_h $. In asymptotic analysis, $\overline{\gamma}_k \rightarrow \infty$ case has been ignored as it indicates a successful wiretapping probability. Hence, the diversity gain is zero for that particular case.

\subsection{Strictly Positive Secrecy Capacity (SPSC) Performance}
The SPSC, which points to the probability of existence of a non-negative secrecy capacity, is a fundamental benchmark in secure communications, mathematically can be expressed as \cite{lei2015performance}
\begin{align}
    \label{abcd9}
    SPSC= \Pr\{ \mathcal{S}_{c} > 0 \}=1 - SOP_L |_{\mathcal{R}_t=0}.
\end{align}
Substituting $\mathcal{R}_t=0$ into \eqref{abcd4}, we can obtain the expression of SPSC.

\subsection{Novelty of this Work}

As AKMS fading channel represents a generalized fading model, it can be used to represent various other classical channels depending on the values of the shape parameters \cite[Table~1]{noor2021on}. Assuming $\alpha_i=\alpha$, $\kappa_i=0$, and $\mu_i=\mu$, the ASC results utilizing \eqref{asc1} perfectly matches with the results of \cite{lei2017secrecy} which represents an $\alpha-\mu$ fading channel. For $\alpha_i=2$, $\kappa_i=\kappa$, $\mu_i=\mu$, and $m_i \rightarrow \infty$, the SOP and SPSC results obtained via \eqref{abcd4} and \eqref{abcd9} completely agree with the corresponding results of a $\kappa-\mu$ fading channel in \cite{bhargav2016secrecy}. Likewise, the results presented in \eqref{abcd4} can be shown to match with \cite[Eq.~(6),]{srinivasan2018secrecy} for $\alpha_i=2$, $\kappa_i=\kappa$, $\mu_i=\mu$, and $m_i=m$. Furthermore, setting $\alpha_i=\alpha$, $\mu_i=2 \mu$, $m_i=\mu$, $\alpha_i=\alpha$, $\kappa_i=\kappa$, $\mu_i=\mu$, and $m_i \rightarrow \infty$, \eqref{abcd4} can be reduced to \cite[Eq.~(9),]{8556482}
and \cite[Eq.~(11),]{8556482}, respectively.

\section{Numerical Results}
In this section, the numerical outcomes utilizing the performance metrics of \eqref{asc1}, \eqref{abcd4}, \eqref{abcd5}, and \eqref{abcd9} are represented and further authenticated via Monte-Carlo simulations by generating an AKMS random variable in MATLAB and averaging $10^{6}$ channel realizations for obtaining each value of $\mathcal{S}_{c}$. As the infinite series converges quickly after few terms, all of them are truncated after the first twenty terms with an accuracy factor of $10^{-4}$.

\begin{figure}[!ht]
\vspace{-45mm}
    \centerline{\includegraphics[width=0.67\textwidth]{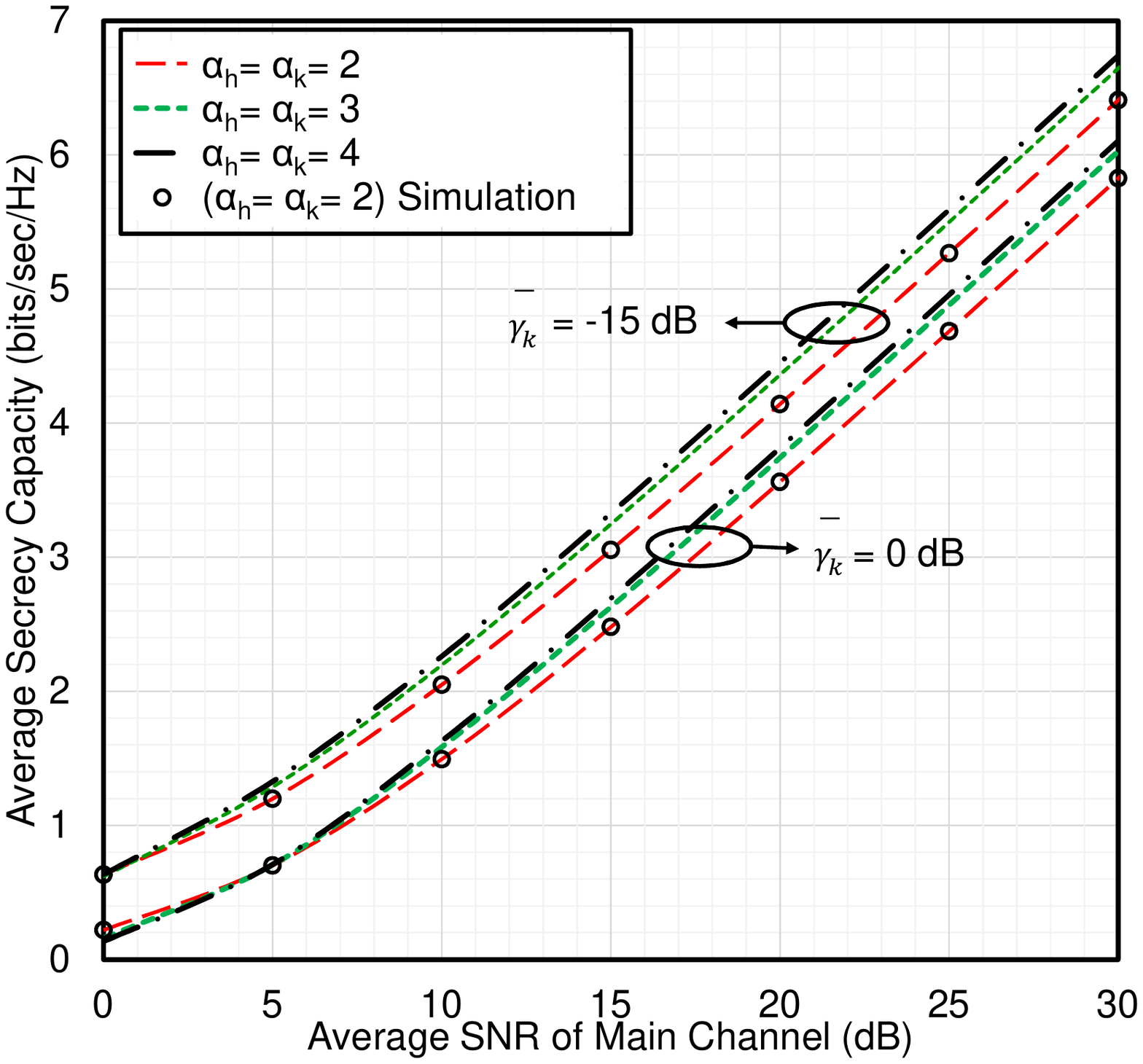}}
    \vspace{-35mm}
    \caption{The ASC versus $\overline{\gamma}_h$ for selected values of $\alpha$ where $m_h=m_k=100$, $\kappa_h=\kappa_k=1$,  and $\mu_h=\mu_k=1$.}
    \label{fig1}
    \vspace{-2mm}
\end{figure}

\begin{figure}[!ht]
\vspace{-25mm}
    \centerline{\includegraphics[width=0.58\textwidth]{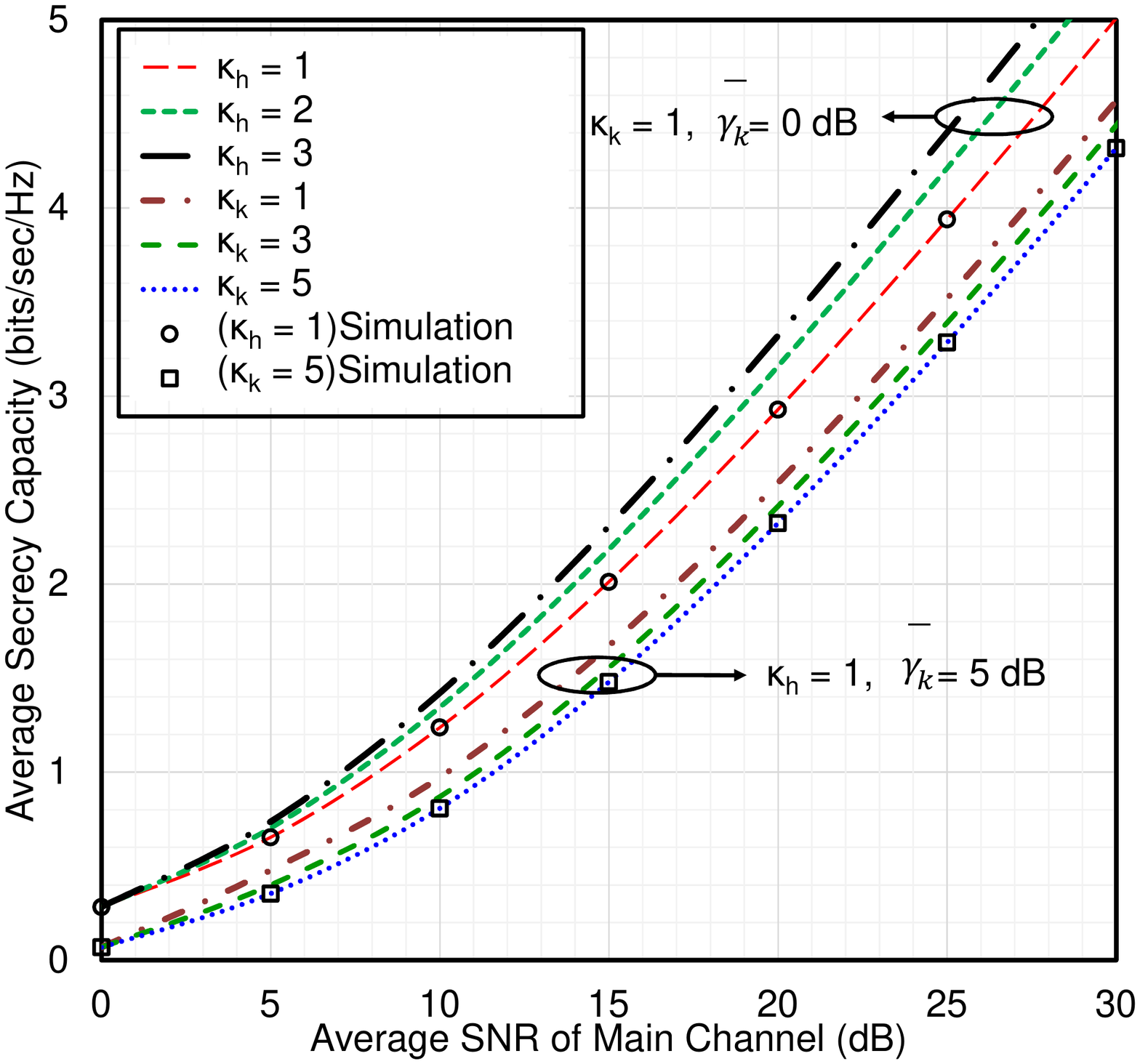}}
    \vspace{-35mm}
    \caption{The ASC versus $\overline{\gamma}_h$ for selected values of $\kappa_h$ and $\kappa_k$ where $m_h=m_k=100$, $\alpha_h=\alpha_k=1$ and $\mu_h=\mu_k=1$.}
    \label{fig4}
    \vspace{-5mm}
\end{figure}

The effects of $\alpha_{i}$ and $\kappa_{i}$ on the ASC are depicted in Figs. \ref{fig1} and \ref{fig4} by plotting ASC against $\overline{\gamma}_h$. It is noted that an increase in $\alpha$ and $\kappa_{h}$ cause a remarkable improvement of ASC whereas the ASC performance is degraded with  $\kappa_{k}$. The reason for this result is that the increase of $\alpha$ and $\kappa_{h}$ improves the $\mathcal{T}-\mathcal{H}$ link but with the increase of $\kappa_{k}$ the $\mathcal{T}-\mathcal{K}$ link is improved. A comparison between the numerical and simulation results discloses that the simulation results are as good as the numerical results that point to the authorizations of the deduced mathematical expressions.

\begin{figure}[!ht]
\vspace{-25mm}
    \centerline{\includegraphics[width=0.58\textwidth]{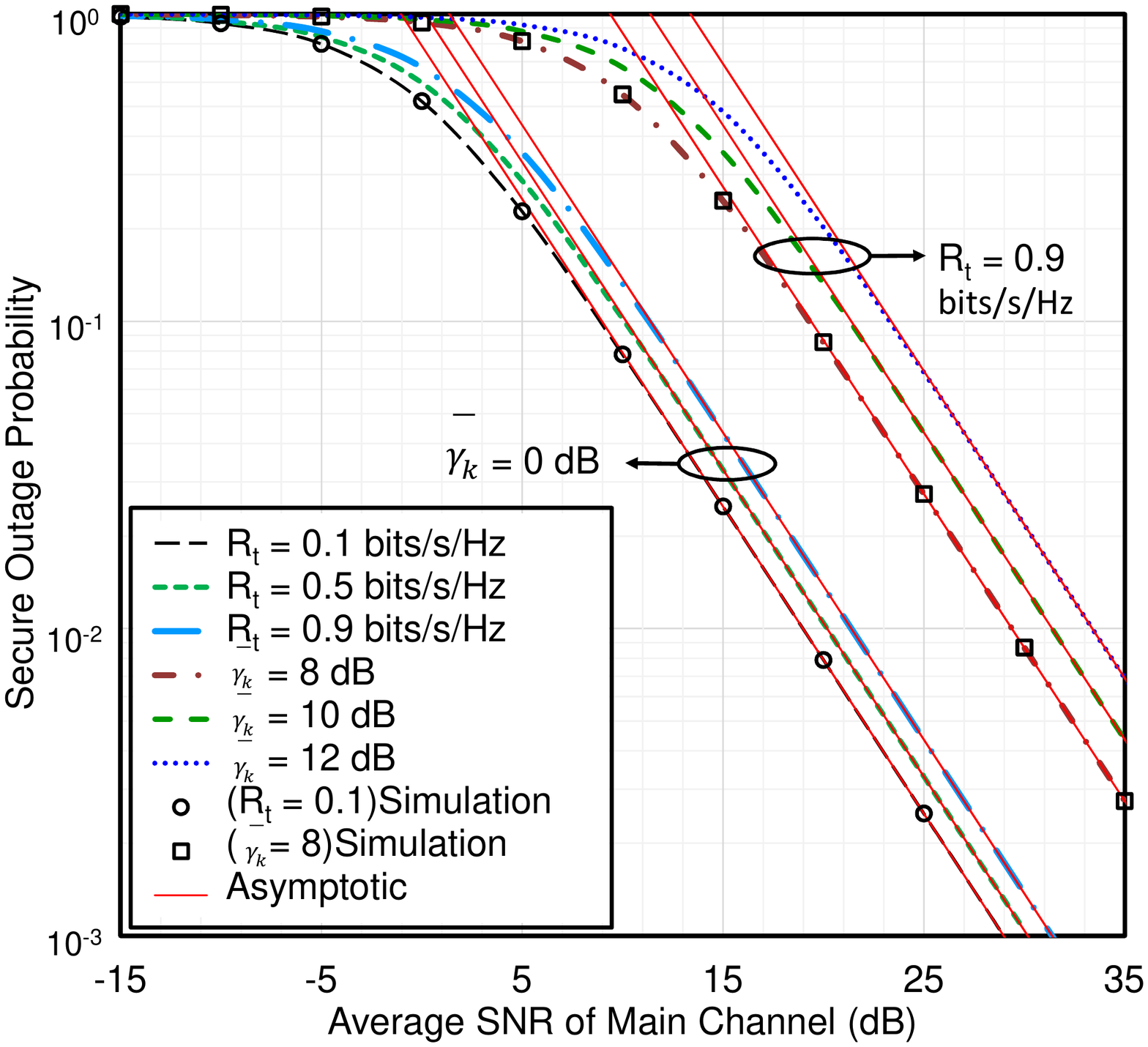}}
    \vspace{-35mm}
    \caption{The SOP versus $\overline{\gamma}_h$ for selected values of $R_s$ and $\overline{\gamma}_k$ where $\alpha=2$, $m_h=m_k=15$, $\kappa_h=\kappa_k=1$,
    and $\mu_h=\mu_k=1$.}
    \label{fig2}
    \vspace{-2mm}
\end{figure}

\begin{figure}[!ht]
\vspace{-40mm}
    \centerline{\includegraphics[width=0.67\textwidth]{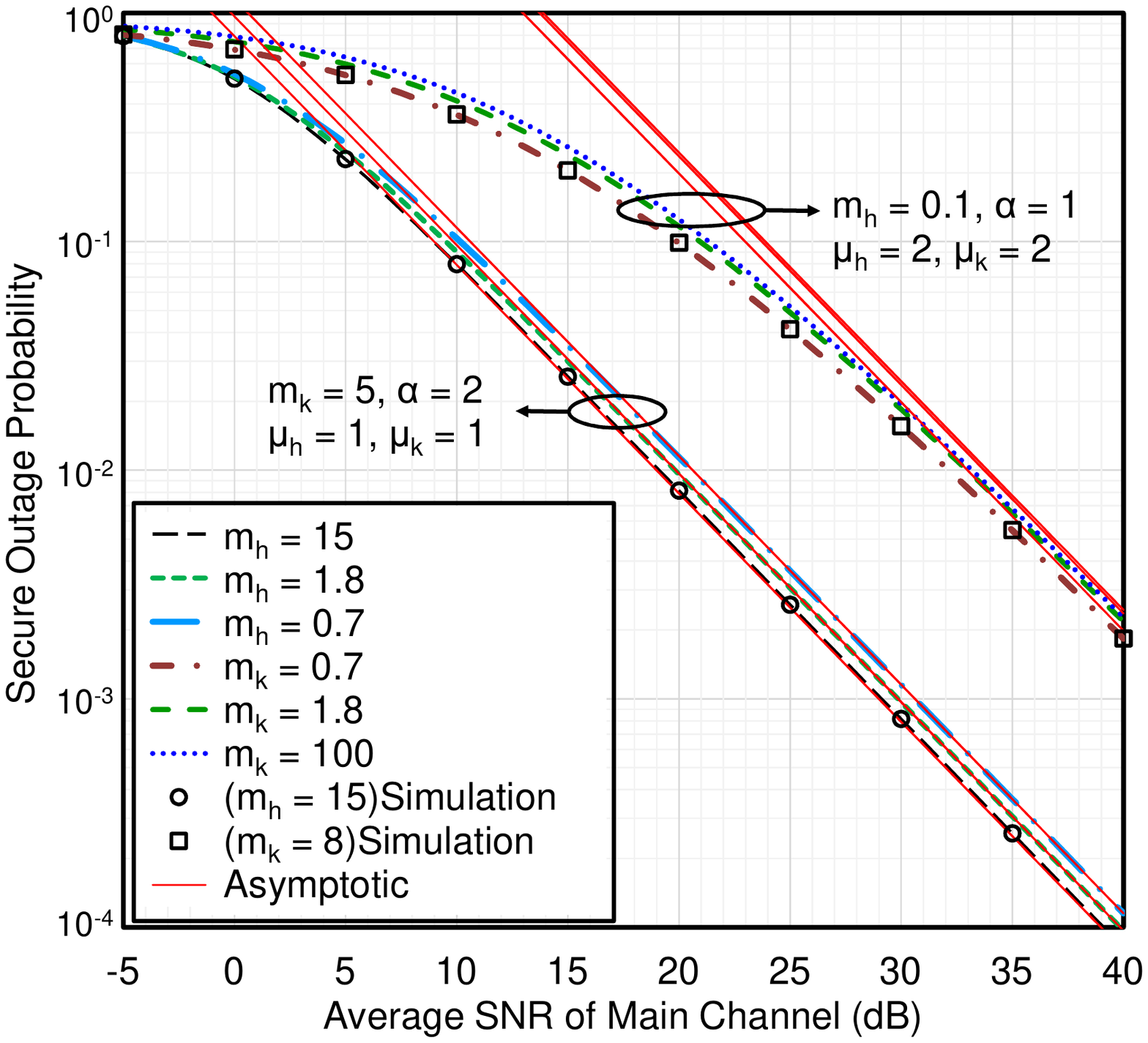}}
    \vspace{-35mm}
    \caption{The SOP versus $\overline{\gamma}_h$ for selected values of $m_h$ and $m_k$ where $\kappa_h=\kappa_k=1$, $\mathcal{R}_{t}=0.5$ bits/s/Hz and $\overline{\gamma}_k = 0$ dB.}
    \label{fig5}
    \vspace{-5mm}
\end{figure}

The SOP is depicted against $\overline{\gamma}_h$ in Figs. \ref{fig2} and \ref{fig5} to observe how $\overline{\gamma}_k$, $\mathcal{R}_{t}$ and $m_{i}$ influence the secure outage characteristics. It is observed from Fig. \ref{fig2} that the SOP gradually increases with both $\overline{\gamma}_k$ and $\mathcal{R}_{t}$. This is because an increase in $\mathcal{R}_{t}$ increases the probability of dropping $\mathcal{S}_{c}$ below $\mathcal{R}_{t}$. On the other hand, an increased $\overline{\gamma}_k$ indicates an enhanced capability of successful wiretapping by the eavesdroppers. In Fig. \ref{fig5}, two cases are considered to demonstrate the effects of shadowing parameters over $\mathcal{T}-\mathcal{H}$ and $\mathcal{T}-\mathcal{K}$ links separately. It is noted that an increase in $m_{h}$ enhances the outage performance. On the contrary, the outage performance is degraded with $m_{k}$. Actually, these results reveal that the overall shadowing of the $\mathcal{T}-\mathcal{H}$ and $\mathcal{T}-\mathcal{K}$ links decreases as the corresponding shadowing parameters increase from 0 to $\infty$. Moreover, it is noted that at a high SNR regime,  the asymptotic curves exactly approach the exact SOP curves.

\begin{figure}[!h]
\vspace{-40mm}
    \centerline{\includegraphics[width=0.67\textwidth]{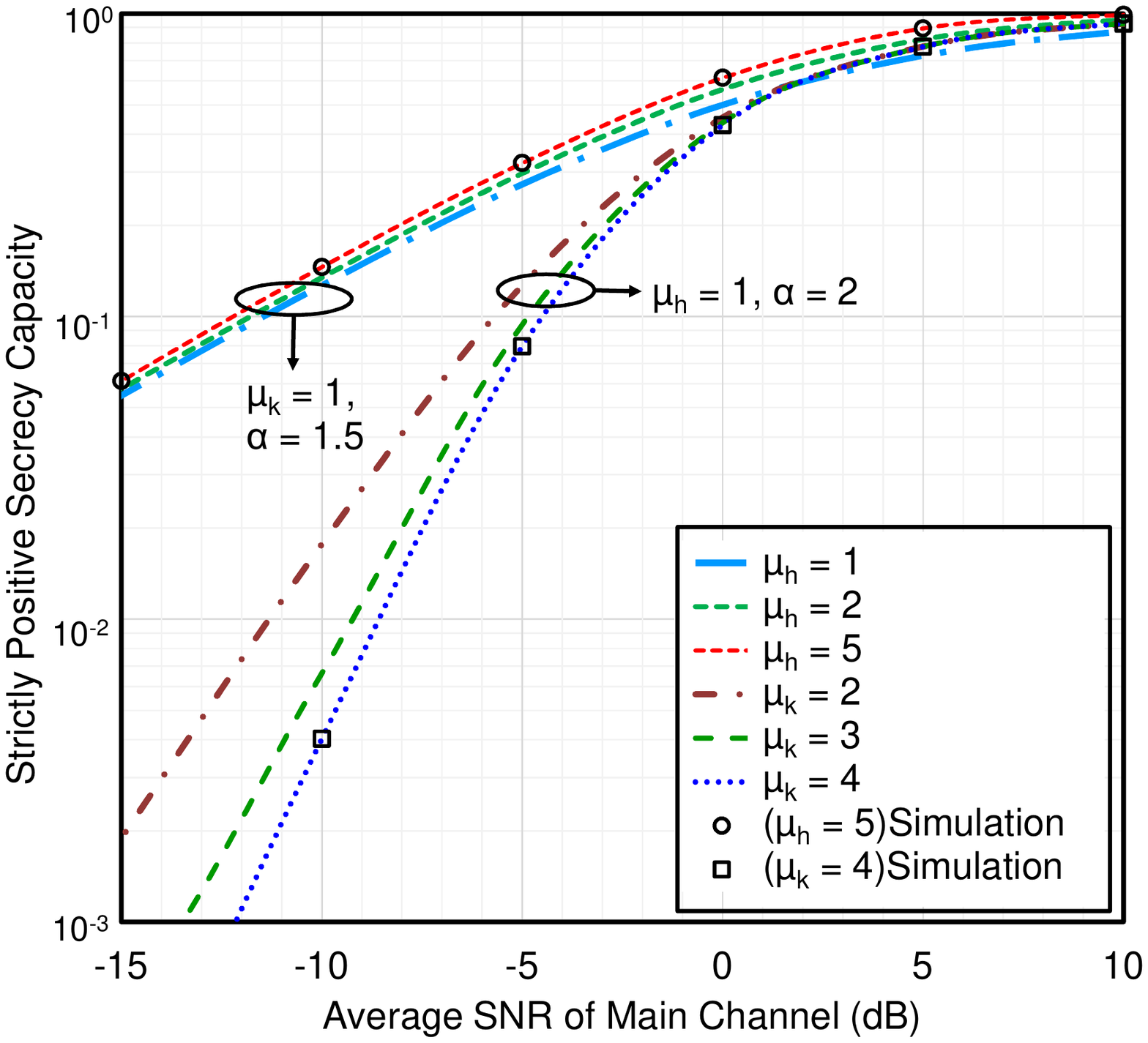}}
    \vspace{-35mm}
    \caption{The SPSC versus $\overline{\gamma}_h$ for selected values of $\mu_h$ and $\mu_k$ where $m_h=m_k=15$, $\kappa_h=\kappa_k=1$ and $\overline{\gamma}_k = 0$ dB.}
    \label{fig3}
    \vspace{-5mm}
\end{figure}

In Fig. \ref{fig3}, the impacts of $\mu_{h}$ and $\mu_{k}$ are demonstrated in terms of SPSC. As an increase in $\mu_{h}$ and $\mu_{k}$ reduces the fading of the corresponding channels, it is clearly observed from the figure that the SPSC improves significantly with $\mu_{h}$ and deteriorates with $\mu_{k}$.

\section{Conclusion}
This paper focuses on the assessment of the secrecy performance of a wireless network over AKMS fading channel wherein a single eavesdropper is trying to overhear and decode the transmitted secret messages. This investigation includes examining the impacts of all system parameters on the secrecy performance via deriving expressions of three secrecy parameters i.e. SPSC, ASC, and SOP. Utilizing these closed-form expressions some numerical outcomes were presented and further authenticated via Monte-Carlo simulations to demonstrate that analytical and simulation results are in close agreement with each other. Additionally, to evaluate diversity gain, asymptotic SOP analyses were derived. It is observed that diversity gain is dependent on $\tilde{\alpha}$ and $\mu_{h}$, but completely independent of the shadowing parameters of $\mathcal{T}-\mathcal{H}$ and $\mathcal{T}-\mathcal{K}$ links. Finally, it can be concluded that the secrecy performance is significantly affected by the fading and shadowing parameters, particularly of the main link rather than the eavesdropper link.

\appendix

\section*{Appendix}
\section{Proof of ASC}
\label{ASC}

The identity in \cite[Eq.~11,]{adamchik1990algorithm} is utilized for representing the exponential and logarithmic components in terms of Meijer’s G function. Utilizing \cite[Eq.~21,]{adamchik1990algorithm}, the term $\Im_1$
is expressed as
\begin{align}
    \nonumber
    &\Im_1=\int_{0}^{\infty} ln(1+\gamma_h) f_h(\gamma_h)F_k(\gamma_h) d\gamma_h
    \\
    \label{asc2}
    &= \sum_{j_h=0}^{\infty} 
    \frac{\sqrt{2}a_h\mathcal{A}_h d_h^{j_h}}{\alpha(2\pi)^{\alpha-.5}  j_h!} \left( I_1 -\sum_{j_k=0}^{\infty} \sum_{l_k=0}^{\mu_k+j_k-1}\frac{\mathcal{B}_k{b_k}^{l_k} }{a_k^{-1}\tilde{\alpha}l_k!}I_2 \right),
\end{align}
where $I_1=\scriptsize{G_{2\alpha,2+2\alpha}^{2+2\alpha,\alpha}\left[\frac{b_h^2}{4}\left|
     \begin{array}{c}
      x_1,x_2\\
      x_5,x_1,x_1\\
     \end{array}
 \right.\right]}$, $I_2=\scriptsize{G_{2\alpha,2+2\alpha}^{2+2\alpha,\alpha}\left[\frac{(b_h+b_k)^2}{4}\left|
     \begin{array}{c}
      x_3,x_4\\
      x_5,x_3,x_3\\
     \end{array}
 \right.\right]}$, 
 $x_1=\triangle(\alpha,-\tilde{\alpha} (\mu_h+j_h))$,
 $x_2=\triangle(\alpha,1-\tilde{\alpha} (\mu_h+j_h))$, 
 $x_3=\triangle(\alpha,-\tilde{\alpha} (\mu_h+j_h+l_k))$,
 $x_4=\triangle(\alpha,1-\tilde{\alpha} (\mu_h+j_h+l_k))$, 
 $x_5=\triangle(2,0)$, $\triangle(x,a)=\frac{a}{x},\frac{a+1}{x},\text{...},\frac{a+x-1}{x}$, and $G_{p,q}^{m,n}\left[\scriptsize{x\biggl |
\begin{array}{c}
 \alpha_{1},....,\alpha_{p} \\
 \beta_{1},....,\beta_{q} \\
\end{array}}
\right]$ signifies the Meijer's $G$ function. Here, an assumption of $\tilde{\alpha}_h=\tilde{\alpha}_k=\tilde{\alpha}$ is taken into consideration during the ASC analysis for the ease of mathematical calculations.
Similarly, $\Im_2$ is expressed as
\begin{align}
    \nonumber
    &\Im_2=\int_{0}^{\infty} ln(1+\gamma_k) f_k(\gamma_k)F_h(\gamma_k) d\gamma_k
    \\
    \label{asc3}
    &= \sum_{j_k=0}^{\infty} 
    \frac{\sqrt{2}a_k\mathcal{A}_k d_k^{j_k}}{\alpha(2\pi)^{\alpha-.5}  j_k!} \left( I_3 -\sum_{j_h=0}^{\infty} \sum_{l_h=0}^{\mu_h+j_h-1}\frac{\mathcal{B}_h{b_h}^{l_h} }{a_h^{-1}\tilde{\alpha}l_h!}I_4 \right),
\end{align}
where $I_3=\scriptsize{G_{2\alpha,2+2\alpha}^{2+2\alpha,\alpha}\left[\frac{b_k^2}{4}\left|
     \begin{array}{c}
      z_1,z_2\\
      x_5,z_1,z_1\\
     \end{array}
 \right.\right]}$,
 $I_4=\scriptsize{G_{2\alpha,2+2\alpha}^{2+2\alpha,\alpha}\left[\frac{(b_h+b_k)^2}{4}\left|
     \begin{array}{c}
      z_3,z_4\\
      x_5,z_3,z_3\\
     \end{array}
 \right.\right]}$,
 $z_1=\triangle(\alpha,-\tilde{\alpha} (\mu_k+j_k))$,
 $z_2=\triangle(\alpha,1-\tilde{\alpha} (\mu_k+j_k))$,
 $z_3=\triangle(\alpha,-\tilde{\alpha} (\mu_k+j_k+l_h))$, and
 $z_4=\triangle(\alpha,1-\tilde{\alpha} (\mu_k+j_k+l_h))$. Again, $\Im_3$ is expressed in a similar way as
 \begin{align}
     \nonumber
    &\Im_3=\int_{0}^{\infty} ln(1+\gamma_k) f_k(\gamma_k)d\gamma_k
    \\
    \label{asc4}
    &= \sum_{j_k=0}^{\infty} 
    \frac{\sqrt{2}a_k\mathcal{A}_k d_k^{j_k}}{\alpha(2\pi)^{\alpha-.5}  j_k!} G_{2\alpha,2+2\alpha}^{2+2\alpha,\alpha}\left[\frac{b_k^2}{4}\left|
    \begin{array}{c}
    z_1,z_2\\
    x_5,z_1,z_1\\
    \end{array}
    \right.\right].
 \end{align}
 
\section{Proof of Asymptotic SOP}
\label{SOP}

We assume $\overline{\gamma}_h \rightarrow \infty$ with fixed $\overline{\gamma}_k$ for which the dominating term in \eqref{abcd1} corresponds to $j = 0$. Utilizing the approximation, $\Upsilon(s, x) \approx x^s/s$ as $x \rightarrow 0$, \eqref{abcd1} can be expressed as
\begin{align}
     \label{abcd7}
     F_{i}(\gamma)=\frac{a_i}{\tilde{\alpha}_i \mu_i} \gamma^{\tilde{\alpha}_i \mu_{i}}.
\end{align}
By substituting \eqref{abcd6} and \eqref{abcd7} into \eqref{abcd3}, the asymptotic expression for lower bound of the SOP is obtained as shown in \eqref{abcd5}
where $a'_k=\frac{{m_{k}}^{m_{k}} \alpha_{k}  }{2 {c_{k}}^{\mu_{k}} \Gamma(\mu_{k}) (\mu_{k} \kappa_{k}+m_{k})^{m_{k}}}$,  $b'_{k} =\frac{1}{c_{k}} $,  $d'_{k}=\frac{\mu_{k} \kappa_{k}}{c_{k} (\mu_{k} \kappa_{k} + m_{k})}$, $a'_h=\frac{{m_{h}}^{m_{h}} \alpha_{h}  }{2 {c_{h}}^{\mu_{h}} \Gamma(\mu_{h}) (\mu_{h} \kappa_{h}+m_{h})^{m_{h}}}$.

\bibliography{sample}

\end{document}